# ChloroScan: Recovering plastid genome bins from metagenomic data

Short title: ChloroScan: plastid MAGs

Yuhao Tong[1], Vanessa Rossetto Marcelino[1,2,3], Robert Turnbull[4], Heroen Verbruggen[1,5].
1. Melbourne Integrative Genomics, School of BioSciences, University of Melbourne, Melbourne, VIC 3010, Australia.
2. Institute of Agrochemistry and Food Technology, Spanish National Research Council (CSIC), Valencia, 46980, Spain.
3. Department of Microbiology and Immunology at the Peter Doherty Institute for Infection and Immunity, University of Melbourne, Melbourne, VIC 3010, Australia
4. Melbourne Data Analytics Platform (MDAP), Melbourne Connect, University of Melbourne, Melbourne, VIC, Australia.
5. CIBIO, Centro de Investigação em Biodiversidade e Recursos Genéticos, InBIO Laboratório Associado, Campus de Vairão, Universidade do Porto, 4485-661 Vairão, Portugal.

Contact: yuhtong@student.unimelb.edu.au

# Abstract

Genome-resolved metagenomics has contributed largely to discovering prokaryotic genomes. When applied to microscopic eukaryotes (protists), challenges such as the high number of introns and repeat regions found in nuclear genomes have hampered the mining and discovery of novel protistan lineages. Organellar genomes are simpler, smaller, have higher abundance than their nuclear counterparts and contain valuable phylogenetic information, but are yet to be widely used to identify new protist lineages from metagenomes. Here we present "ChloroScan", a new bioinformatics pipeline to extract eukaryotic plastid genomes from metagenomes. It incorporates a deep learning contig classifier to identify putative plastid contigs and an automated binning module to recover bins with guidance from a curated marker gene database. Additionally, ChloroScan summarizes the results in different user-friendly formats, including annotated coding sequences and proteins for each bin. We show that ChloroScan recovers more high-quality plastid bins than MetaBAT2 for simulated metagenomes. The practical utility of ChloroScan is illustrated by recovering 16 medium to high-quality metagenome assembled genomes (MAGs) from four protist-size fractioned metagenomes, with several bins showing high taxonomic novelty. The ChloroScan code is available at https://github.com/Andyargueasae/chloroscan under Apache-2.0 license.

# 1. Introduction

The amount of sequenced data for microbiomes has ballooned in the last decade [1]. Genome-resolved metagenomics (GRM) became a widely used approach to analyze these data for environmental microbiomes, offering numerous insights of their evolution, ecology and diversity [2]. It incorporates *de novo* assembly to build contigs and binning algorithms to cluster these into metagenome-assembled genomes (MAGs). These MAGs can then be used in phylogenetic analyses [3], or to understand microbial metabolism, adaptations to different environments [4], or to gain insights into the genomics of biological interactions [5], [6].

GRM has greatly expanded our knowledge of prokaryotes. For example, it has led to the discovery of nanosized Candidate Phyla Radiation (CPR) representing at least 15% of bacterial phyla [7], and the discovery of Asgard Archaea that drastically altered our understanding of eukaryote origins [8], [9], [10]. However, metagenomics has not yet achieved similar breakthroughs for microbial eukaryotes (protists) [11]. Like prokaryotes, microbial eukaryotes have significant roles in different environments [12],

[13], [14] and their genomes are valuable resources to understand their physiology, evolution and ecological niche [15].

Currently, only a minor fraction of eukaryotes has had their genomes sequenced [16], [17]. It was hoped that metagenomics may help solve this data shortage by taking advantage of any eukaryotic data that may be present in large metagenome libraries like Tara Oceans [18], but some methodological issues and the much more complex eukaryotic genome features have stood in the way of fully achieving this [11]. First, abundant repeats, non-coding regions and large genome sizes in eukaryotes are challenging for current bioinformatic tools [11], [19]. Second, eukaryotes are usually less abundant than prokaryotes [20]. Third, most existing bioinformatics tools are geared towards prokaryotic data [19]. For example, the reference databases for marker gene annotations (e.g.: CheckM [21] and CheckM2 [22]) or taxonomy classifications (e.g.: GTDB-tk [23]) are specific to prokaryotes [22], [23], [24], [25], [26]. Despite these hindrances, recent work has recovered hundreds of eukaryotic MAGs from shotgun metagenomes [19], [27], [28], [29]. The taxa for which genomes have been recovered, however, are often those with small genomes like prasinophytes [19], [29], and any MAGs recovered from other eukaryotic taxa are largely incomplete [30], hindering comparative genomic analyses [29].

One potential path forward may be to focus on the organellar (plastid and mitochondrial) genomes of eukaryotes. Plastids, for example, are remnants of the ancestral cyanobacteria involved in the ancient endosymbiosis events that gave rise to the organelles [31]. Through endosymbiotic gene transfer (EGT) and gene loss, the gene content of plastid genomes has become highly streamlined [31], making them significantly smaller than many prokaryote genomes. Usually, they have a higher copy number, and hence higher sequence coverage, when compared to nuclear genomes [32], [33]. They also commonly have a distinct nucleotide composition [34], which may facilitate their assembly and binning into MAGs.

Organellar genes are frequent targets for phylogenetic analysis, with plastid genes such as *rbcL and tufA* frequently used in species delimitation and DNA barcoding [35], [36], and plastid genome-scale trees are offering great advances to our knowledge of algal evolution [37], [38], [39]. Currently, ~2,300 complete plastid genomes from ca. 1100 algal species are recorded in GenBank [40], a small fraction of the over 50,000 algal species estimated to exist by 2024 [41]. Metagenome-derived plastid MAGs (ptMAGs) could help mitigate the data sparsity and facilitate studying their evolution and genomic features. Early attempts successfully recovered plastid genomes from Sargasso Sea metagenomes with high sequence coverage [32]. Such motivation already led to bioinformatic developments, including plastiC [42]: the only workflow targeting metagenomic plastid sequences, by clustering contigs from metagenomes via MetaBAT2 [43] and keeping putative plastid bins (predicted by Tiara [44]) without prior filtering of assemblies. Another team proposed a GRM pipeline incorporating human-guided binning from anvi'o workflow (manually grouping contigs into bins based on k-mer and coverage) [45], which does not easily scale to larger metagenome datasets [46]. The results from these recent attempts offer valuable data on the feasibility of extracting plastid MAGs from metagenomic data. However, handling large volumes of data in metagenomics requires scalable workflows and automation, and a plastid genome-specific database of marker genes to assess MAG quality [21] and guide binning [47] is missing in existing approaches.

In this study, we present ChloroScan, an automated computational workflow that targets plastid genomes in metagenomes. It addresses the above challenges by automating targeted binning, taxonomic prediction and quality control, guided by a manually curated database of plastid genes. ChloroScan generates information-rich summaries of the obtained ptMAGs to help interpret and use them in downstream applications. We benchmark ChloroScan against currently used tools using simulated and real metagenome data.

## 2. Materials and Methods

### 2.1 ChloroScan workflow overview

ChloroScan is a Snakemake-based [48] workflow with the command line interface wrapped by snk [49], to infer ptMAGs from metagenome contigs, with the utilities mostly written in Python and Unix bash. It is underpinned by a deep learning-based module to predict contigs of plastid origin and a manually curated database to guide metagenome binning and quality assessment (Figure 1).

The first step in ChloroScan classifies contig categories using a deep learning contig classifier (Corgi [50]) to denoise the assembly and offer a plastid-enriched assembly for downstream processes. For each contig, Corgi gives a probability of the target sequence being one of five RefSeq categories: eukaryotic nuclear, mitochondrial, plastidic, bacterial, archaeal and unknown otherwise. If the probability of one sequence type exceeds all others and reaches a given threshold, the sequence is assigned to that category. By default, ChloroScan runs Corgi with a probability threshold of ≥0.50 for the plastid category, and retaining only contigs ≥ 1000bp, but these settings can be altered to optimize data mining performance in different data contexts.

The second step: metagenome binning is a key step of the ChloroScan workflow. We employ binny [47], which uses k-mer frequencies and sequence abundance in clustering contigs and uses marker genes to guide the clustering of contigs into bins. Binny's original database to guide clustering focuses on prokaryotes (CheckM [21]). Hence, we designed a custom database for plastid marker genes following its rationale [21] (Supplementary Materials), transforming it to recover plastid bins. We altered binny to not run read depth calculations, but rather do this more efficiently within the core ChloroScan workflow. For the recovery of single-contig MAGs (scMAGs), we set default thresholds of 30 for marker gene count and 85% for marker completeness.

Following binning, taxonomic classification adds an extra layer of information for users to interpret their data by showing contig-level taxonomy identification. ChloroScan uses CAT/BAT [51] with default settings to predict the taxa of contigs based on the ORFs it predicts and searches the ORFs against a protein database to infer the likely taxonomic origin of contigs. For this step, we designed a custom protein reference database that includes the uniref90 [52] database to provide taxonomic breadth along with an extensive custom database of plastid-encoded protein sequences of algae to enhance the resolution of taxonomic identifications. This dataset replaces the original CAT/BAT-prepared nonredundant protein database (nr) which requires more download and storage capacity, runtime, and memory usage.

After these computations, ChloroScan uses a range of custom Python scripts to deliver diverse user-friendly outputs and summaries that enable users to make sense of their data and use the results for a range of downstream applications. The first step in this section is to detect contamination in bins. Due to the lack of published refinement tools for ptMAGs, we adopted a conservative contamination detection based on contig-level taxonomy prediction by CAT/BAT to highlight bins containing contigs with ambiguous or non-eukaryotic taxon predictions and without ORFs matching our plastid marker gene database. We also provide community-level taxonomic information of the Corgi-filtered metagenome with Kronatools [53]. ChloroScan produces a range of graphs, including a scatterplot of GC contents vs. Log10 transformed average coverage) showing the homogeneity of bins, pie charts showing the taxonomic assignments of the contigs in each bin and a violin plot demonstrating the contig coverage distribution for each bin. A spreadsheet reports detailed contig statistics, including their length, marker genes and taxon prediction.

Finally, ChloroScan predicts the coding nucleotide sequences and proteins of bins. Due to the potential occurrence of gene fragments in complex metagenome assemblies, we chose FragGeneScanRs [54]. The coding sequences and proteins of each bin are provided in FASTA format, which can directly feed into downstream comparative and/or phylogenomic analyses. The location of genes on the contigs is also given as a GFF3 file.

## 2.2 Benchmarking ChloroScan with synthetic data

ChloroScan was benchmarked using simulated marine metagenomic datasets. We designed simulated samples containing prokaryotic, eukaryotic nuclear, eukaryotic mitochondrial, and plastid genomes. We generated two simulated samples with different sets of relative abundances of each genome using CAMISIM [55], [56], with the N50 for sample to be 17,762 bp and sample 2 to be 15,843 bp; the simulation method is described in supplementary materials. These samples contain 12 plastid genomes: 11 canonical plastid genomes and one dinoflagellate plastid genome characterized by minicircles. We generated read depth profiles by aligning the simulated reads to the metagenome assemblies using minimap2 (version 2.26) [57]. As we failed to run plastiC (Supplementary materials), we used its underlying binner (MetaBAT2 [43]) for benchmarking comparison. When running ChloroScan, we set the contig length cutoff to 1500 bp, to match MetaBAT2's minimum contig length cutoff. Other settings remained default, and we set the default minimum completeness of bins to 50% to recover as many bins as possible despite their sizes. Binning results from these tools and ground truth were analyzed using AMBER [58], including calculations of the F1 score (a summary performance metric combining bin average completeness and purity), bin purity and completeness and accuracy at base pair level (see supplementary materials for definitions). Sequence coverage is known to impact genome recovery. To help exploring the relationship between ChloroScan's recovery outcome and source genomes' coverage, we generated mean read depth for 11 canonical source genomes via mapping simulated reads to these genomes using minimap2 and running depth calculation steps from binny's workflow. To assist in checking the relative abundance of each genome, we computed them via coverm's genome pipeline [59]. The corresponding data are in supplementary file S1.

## 2.3 Application to real metagenomes

We then applied ChloroScan to four marine metagenomes from the Tara Oceans expedition datasets, using the three size fractions under the category "protist": SAMEA2189670 (0.8–5μm), SAMEA2732360 (180–2,000 μm), SAMEA2657032 and SAMEA2732613 (20–180 μm). Most samples were collected from surface sea waters, except SAMEA2732613 at 40m depth. We downloaded their MEGAHIT assemblies from SPIRE [60], consisting of ~24M contigs. We then mapped the samples' raw reads using minimap2 (version 2.26) in short-read mode. Due to a lack of prior knowledge of these samples, to assess those medium and high-quality ptMAGs with more confidence of their homogeneity, we chose a conservative set of parameters while running ChloroScan, by setting minimum completeness and purity cutoff to 70% and 90% [61]. Other settings remained as default. Later, to inspect the resulting bins' taxonomy, we retrieved their *rbcL* genes via orthofisher [62] (Supplementary materials) and blasted them against nonredundant protein database.

# 3. Results

## 3.1 ChloroScan recovers high-quality ptMAGs from synthetic metagenomes

To compare ChloroScan's performance against that of similar software, we benchmarked it alongside MetaBAT2, the binner used by plastiC [42]. Both solutions produced plastid MAGs (Figure 2), but we found that those produced with ChloroScan had higher overall quality and purity, with the average F1 score and accuracy at base pair level for the two simulated samples 27.5% and 23.3% higher than MetaBAT2, respectively (Table S1 and S2). In total, ChloroScan recovered 8 and 9 bins from these two synthetic metagenomes, with 6 and 8 of these high-quality according to MIMAG [61]. Five and six were near-complete single-contig MAGs. MetaBAT2 recovered 6 and 7 bins, of which only 3 were considered high quality [61] in both samples (Figure 2c, d). For sample 1, the remaining plastids were also retrieved as fragmented genome sequences by ChloroScan, with completeness ranging from 2.2% to 94.5% (see supplementary file S1), resulting in an overall accuracy of 0.848 and an F1 score of 0.861. MetaBAT2's accuracy was 0.753 and F1 score of 0.573 for this sample (Table S1). For sample 2, the plastid genomic sequences are less fragmented than in sample 1: the completeness ranges from 73.2% to 99.9% (see supplementary file S1). ChloroScan's accuracy (0.878), F1 (0.735) and average purity per bp (0.982) compared favorably to MetaBAT2's corresponding values (Table S2).

For sample 1, both tools had a similar adjusted random index (ARI): a summary metric to measure the performance of clustering, but ChloroScan recovered bins from a higher percentage of binned nucleotides than MetaBAT2 (Figure S1a). For sample 2, ChloroScan's ARI substantially exceeded MetaBAT2's (Figure S1b). The detailed bin taxonomic compositions show that MetaBAT2 joined the plastid genomes of two *Chlamydomonas reinhardtii* strains (meta1) and of two distinct heterokont species: *Thalassiosira pseudonana* and *Aureococcus anophagefferens* (meta3; Figure 2c and d; Figure 3b). Except those single-contig MAGs (with prefix "sc" in Figure 2 a and b), the multi-contig ptMAGs from sample 1 show wide variation of GC contents, with the ChloroScan-recovered bin 2 having coverage around 30× and the bin 1 having coverage around 3× with more variations in GC contents. In sample 2, the clusters appear coherently in the scatterplot, with nearly uniform read depths within MAGs and less GC content variation compared to synthetic metagenomic sample 1 (Figure S2a and S2b). Neither sample recovered dinoflagellate bins consisting of minicircular chromosomes, and the Corgi-filtered assemblies from two samples did not contain their minicircular chromosomes (Figure S3).

To better understand the average depth threshold above which MAGs can be recovered, we produced the average read depth of source genomes and generated scatterplots to visualize relationships between average depth and fragmentation (quantified as the count of contigs) of the 11 canonical source genomes (Figure 2e and f). We found that in sample 1, the five unbinned genomes have sequence mean coverage of less than 5×. The assembly of *Micromonas commoda* consists of only 8 contigs and only covers a small fraction of the complete genome, so it is the assembly that failed to recover its genome. The other four unbinned genomes' contigs cover a larger fraction of the genome, but with a high level of fragmentation. One exception is *Chlamydomonas reinhardtii* ("chl0"). Despite the high level of fragmentation, it has coverage around 30× and was still recovered with desired completeness and purity. In contrast to sample 1, the quality of binning from sample 2 is significantly better than sample 1, with overall higher coverages across all sampled genomes. Only two source genomes (*Isochrysis galbana* and *M. commoda*) were not recovered, both having a fragmented assembly and a coverage less than 5×.

Importantly, ChloroScan bins have higher average completeness than MetaBAT2 bins (Figure S1c and S1d), and the majority of ChloroScan bins were nearly complete and without any contamination (Figure 2a and b). One important exception is the highly multimeric bin (chl1) produced by ChloroScan for sample 1 (Figure 2a; Figure 3a), which led to the average purity score for ChloroScan bins being lower than that for MetaBAT2 (Figure S1c). This multimeric bin is made up of short contigs from different plastid, nuclear and prokaryote genomes with overall low coverage around 3 × and diverse GC contents (Figure S2a) and was clearly indicated as a taxonomic mosaic based on the contig identifications provided by ChloroScan.

## 3.2 16 ptMAGs recovered from four ocean metagenome samples

ChloroScan's ability to recover plastid genomes from real datasets was clearly illustrated by its application to four Tara Oceans marine metagenomes. Out of ~24 million contigs, ChloroScan classified 16,556 putative plastid contigs and the binning module inferred 16 ptMAGs with completeness > 70% and purity > 90% (medium to high quality according to MIMAG [61]). The scatterplots of two samples (SAMEA2189670 and SAMEA2732360) demonstrate that most bins have homogeneous coverage and slight variation in GC content (Figure 4a; Figure S4).

To demonstrate ChloroScan's ability to recover novel MAGs, we looked at taxonomic predictions for eight bins from the sample SAMEA2732360 (collected from Antarctic coast) and the two bins from SAMEA2189670 (collected from southern Mediterranean), based on the CAT taxonomic predictions implemented in ChloroScan (Figure 4b) and blastn search of the marker gene *rbcL* against core nonredundant nucleotide database (nt). Overall, diatom species are prevalent in SAMEA2732360. Bins 1-6 have almost nesting taxonomic composition predicted by CAT, with contigs either unclassified, belonging to the SAR supergroup: the group that diatoms belong to, or with more specific identity. But there are also exceptions. Bin 0 contains some putative haptophyte contigs in addition to putative SAR contigs, potentially representing slight chimerism. Bin 7 is not classified based on our protein database containing all protistan plastid proteins and Uniref90 proteins. Our blastn results using the *rbcL* gene recovered from each of these bins provided more fine-grained taxonomic identifications. For Bin 0, we got strong hits to several diatom species mainly from Coscinodiscophyceae (supplementary data 1). For Bin 1, the highest hit (94.84%) is with *Fragilariopsis kerguelensis*—a pennate diatom native to Southern Ocean with one of the highest abundances in the sediments [63] (supplementary data 2). Considering the <95% similarity, it seems more likely our bin represents another species within the genus. Bin 2 resembles *Chaetoceros danicus* and *Conticribra weissflogii,* both with 98.29% *rbcL* similarity (supplementary data 3). *Pseudo-nitzschia turgiduloides*—a pennate diatom, matches the *rbcL* sequence of bin 3 at 99.52% identity (supplementary data 4). Bin 4 shows the best hit (95.11%) with the diatom *Eucampia zodiacus,* a species known to cause harmful algal blooms and widely distributed in non-polar waters [64] (supplementary data 5). Our work likely identified a close relative of it that exists in polar water. Bin 5's identity is narrowed to *Chaetoceros*, highly like *Chaetoceros gelidus* with 99.93% similarity (supplementary data 6). Bin 6's *rbcL* had *Corethron hystrix* as the top blastn hit (93.55%), indicating relatedness but not that exact species (supplementary data 7). It has the highest abundance among all bins (Figure 4a). Finally, bin 7's *rbcL* is identical to a eudicot (muskmelon), which is likely a sample/lab contaminant (supplementary data 8). In sample SAMEA2189670, the bin 0 contigs has a nested SAR origin predicted by CAT. The *rbcL* blast deduces it as *Pseudo-nitzschia cuspidata* from the family Bacillariaceae (class Bacillariophyceae), with similarity ~97% and query cover 96%. Other taxa in top hits also belong to the same class, giving a putative Bacillariaceae origin (supplementary data 9). Finally, the single-contig MAG (with length 81,223 bp) bin 1 from SAMEA2189670 has predicted proteins resemble that of ochrophytes. The BLAST search does not give close hits, rather 85-87% similarity with several distantly related freshwater chrysophytes (supplementary data 10), suggesting that this bin is likely a new deep-branching lineage within Ochrophyta.

The krona plots (Figure S5) demonstrate that Stramenopiles and pico-sized green algae (e.g. *Chloropicaceae*) dominate the samples, with Bacillariophyta (diatoms) and Ochrophytes dominating. Haptophytes are also found in SAMEA2189670 with the smallest size fraction. Some rare taxa such as Cryptophytes, Discoba and the unicellular red algae family *Galdieriaceae* also exist in the smaller size fraction. In addition, Corgi included a fraction of non-photosynthetic metazoan sequences into the putative plastid contig assemblies for all samples. Overall, most contigs with taxon assigned are restricted to higher ranks (e.g. phylum, class and order) and a considerable fraction of contigs were not assigned any taxon.

## 4. Discussion

We developed ChloroScan, a metagenomic binning workflow targeting plastid genomes, and showed its performance using synthetic and real metagenomes. ChloroScan leverages an existing binning framework designed for prokaryotes [47], but we enhanced its performance for plastid genome binning with a manually designed plastid-encoded marker gene database and settings fine-tuned to retrieve plastid genomes. The utility of ChloroScan to recover plastid genomes from real metagenomes was illustrated by recovering 16 ptMAGs from just four marine metagenome samples (Figure 4; Figure S4).

Our work offers several innovations to the field. Previous works often used manual binning to sort plastid contigs [65], while our automated workflow offers potential for screening much larger datasets. An important innovation in ChloroScan is that it uses plastid marker genes to improve plastid MAG recovery. This approach was known to work well in recovering prokaryotic genomes [47], and our results extend this to plastid genomes (Figure 2a and Figure 3b). To sort out low-quality bins, ChloroScan's visual reports allow the user to easily identify bin quality and possible chimeras as a primary sanity check. An additional novelty we present is incorporating organelle genomes in metagenomic simulations. Current simulated metagenomes are often prokaryote oriented, with only a few examples containing eukaryote genomes [66]. Our approach went well beyond that, with extensive sampling of eukaryote mitochondrial, plastid and nuclear genomes.

Our simulations show that low coverage is a logical limitation that can severely impact on the length of contigs, resulting in them being filtered out prior to binning, and missing the chance of recovering those rarer plastids (Figure 2e and f). The boundary appears to be at ca. 5× average coverage, with source genomes sequenced at lower coverage being more fragmented and less likely to be recovered with high purity, as exemplified by *Micromonas commoda* (Figure 2e and f). Meanwhile, we found that in sample 1 four highly fragmented genomes were not binned despite being assembled at higher coverage and better completeness than *M. commoda*. Likely due to the contig length cutoff of ChloroScan used in this study (1000bp), few contigs from these source genomes entered the binning stage, thus binny failed to include them in binning despite their contigs covering nearly the whole source genomes. Binny was shown to outperform MetaBAT2 in recovering fragmented prokaryotic MAGs [47]. However, plastid genomes are shorter, and the presence of highly fragmented, low-coverage genomes would produce numerous contigs with biased nucleotide frequencies, making downstream binning more vulnerable to producing chimeras (figure 2a).

The taxonomic predictions of ptMAGs recovered from four real marine metagenomes (Figure 4; Figure S4) is consistent with the knowledge from 16S plastid rRNA sequencing analysis that Chlorophyta, Haptophytes and diatoms are some of the most abundant phytoplankton groups in the Tara Oceans samples [67], [68], [69]. Many of these are small and they appear to have substantial biodiversity that is yet to be discovered [67]. For example, prasinophytes, which we found in several samples (Figure S5), are dominant green algal taxa in surface oceans, with prasinophyte clade VII particularly taking high abundance in metabarcoding analysis of Tara Oceans data [70].

Our taxonomic identifications using CAT/BAT and blastn searches of the *rbcL* gene showed: 1) clear evidence for the recovery of plastid genome bins of common ocean phytoplankton, and 2) plastid genomes from novel lineages. At genome-level, bins recovered with taxonomic identification from the polar sample SAMEA2732360 represent the commonly found species from marine microbiomes: diatoms and other ochrophytes, both from the SAR supergroup known to be dominant in Tara Oceans samples [68]. Among recovered MAGs, we recovered a polar-representative MAG: bin 1 that resembles *Fragilariopsis kerguelensis*, which presents with high abundance in Antarctic waters [63]. We also found ptMAGs of cosmopolitan taxa (e.g. *Pseudo-nitzschia* and *Chaetoceros*). Bin 5 from SAMEA2732360, has high *rbcL* similarity (> 95%) to *Eucampia zodiacus* that is found worldwide except in polar waters. Hence, we substantially expand the knowledge regarding this species while offering candidates for future genomic comparisons. Additionally, we recovered an ochrophyte MAG that might from a novel lineage. Bin 1 from SAMEA2189670 shows extremely high contiguity and has only distant matches (ca. 85% identity) among its top ten blastn hits of the *rbcL* gene, with hits of similar identity to several chrysophyte genera (supplementary data 10). Seeing that chrysophytes are a nearly exclusively freshwater lineage [71], we consider this lineage to likely be a yet-to-be-discovered marine ochrophyte lineage (see also [72]). We consider the top blastn hit of the *rbcL* sequence (annotated as

"uncultured bacterium" in NCBI) to be a misidentified algal plastid sequence containing *rbcL*.

Our results also show that ChloroScan did not recover dinoflagellate plastid bins (Figure S3) despite having high overall relative abundance (supplementary file S1). Dinoflagellates have an unusual plastid structure, consisting of several minicircles with each containing a single gene [73]. In the simulated datasets, the contigs corresponding to these minicircles were excluded from the filtered assembly by Corgi despite having high coverage and contiguity. Corgi was trained on RefSeq genomic sequences to predict contig categories, and of the handful of dinoflagellate mini-circle plastid chromosomes currently known, none were found in RefSeq's plastid genome category [74]. Hence, Corgi assigns dinoflagellate plastid sequences relatively low probability of being in the "plastid" category, but this situation could arguably be rectified in future by refining the curation of the training data for Corgi. Previous 16s rRNA sequencing results [68], [69] show that dinoflagellate plastid 16s rRNA has very low relative abundance in real metagenomes. This may indicate that their plastid genomes may also be sequenced at low coverage, lowering the probabilities of successful recovery of their ptMAGs from real metagenomes, but it should also be considered that dinoflagellate 16S rRNA are highly divergent [68] that it may hinder their detections in the survey. Other than dinoflagellates, some other protist lineages (e.g.: apicomplexans) and siphonocladous algae, also have highly reduced plastid genomes [75], [76], which our current workflow is not currently tuned for but offers opportunities for future development.

The Corgi contig classification module inevitably retains some contamination in the filtered assembly (Figure S3), causing potential downstream prokaryotic contamination for the binning module to contend with, as shown in our benchmarking data. The taxonomic reports from CAT/BAT and visualizations offer a chance for users to inspect and remove such contaminant contigs. Our workflow reports contaminant contigs without marker genes or non-eukaryotic predicted identity, which can help users refine their ptMAGs.

Plastid genomes are mainly present in photosynthetic protists (some have reduced or lost plastid genomes like apicomplexans); hence this approach is bound to miss heterotrophic microeukaryotes. However, analogous approaches using mitochondrial genomes might be proposed. Insect studies [77] suggest that this is feasible, and a survey from 2019 revealed substantial mitochondrial genome diversity of flagellated protists from Pacific Ocean samples [78], making it an appropriate target to sample undiscovered heterotrophic protists. ChloroScan is flexible to adapt to mitochondrial MAG recovery, as Corgi also predicts mitochondrial contigs, but a curated database of mitochondrial marker genes from each lineage would need to be designed and implemented to achieve this.

In summary, the ChloroScan workflow facilitates automated binning of ptMAGs from filtered contigs, with the use of a marker gene database to improve binning sensitivity and accuracy. It accurately predicts plastid contigs and recovers ptMAGs with higher completeness and purity from synthetic and real metagenomes. As the availability of reference plastid genomes continues to grow, the marker gene database and corgi training data will capture a broader range of taxa and enhance the precision of binning and taxonomic identifications. A growing number of studies attempting to recover ptMAGs from metagenomes [65], the resulting MAGs can serve as valuable resources to address a variety of downstream biological questions and further bioinformatic advances.

## Acknowledgements

This research was supported by The University of Melbourne's Research Computing Services and the Petascale Campus Initiative. HV is supported by a fellowship from the Fundação para a Ciência e a Tecnologia (CEECIND:2023.06155). VRM was supported by the Australian Research Council (DE220100965) and the Spanish Ministry of Science and Innovation (RYC2023-042907-I).

# Data Accessibility and Benefit Sharing

## Data Accessibility Statement

ChloroScan is available in GitHub: https://github.com/Andyargueasae/chloroscan. Beginners' guide and utilities of ChloroScan can be found at https://andyargueasae.github.io/chloroscan/. The most updated release is v0.1.5 with github link: https://github.com/Andyargueasae/chloroscan/tree/release_v0.1.3. The CAMISIM used in this study can be found at https://github.com/CAMI-challenge/CAMISIM/tree/dev. The CAT database used in the current version of ChloroScan is deposited here in figshare: https://doi.org/10.26188/27990278.

The assemblies and bam files of two synthetic samples used in this study are available in figshare https://doi.org/10.26188/28748540. Real metagenomes' raw reads (SAMEA2189670, SAMEA2732360, SAMEA2657032 and SAMEA2732613) are available in ENA (https://www.ebi.ac.uk/ena/browser/home). Their assemblies assembled by spire workflow can be downloaded in the SPIRE database (https://spire.embl.de/study/TARA_Oceans_protists_metaG?page=1), under the study name "TARA_Oceans_protists_metaG".

The OrthoFinder results on picked genomes, the original and intermediate data to generate the marker gene database, the outputs from ChloroScan for all benchmarking and real metagenomes, the orthofisher output directory for *rbcL* hits from bins from SAMEA2732360 and SAMEA2189670 and the nucleotide coding sequences of these bins are deposited in figshare: https://doi.org/10.26188/28722788. Detailed instructions and codes to construct the plastid marker gene database for binny, and the bash codes used for the analyses in this study are saved in the GitHub repository: https://github.com/Andyargueasae/ChloroScan_reproducibility.git.

## Benefit Sharing

The study has no benefits to report.

# Authors' Contributions

Yuhao Tong contributed to workflow construction and adjustment, metagenome data simulation, GitHub repository maintenance and data analysis. Robert Turnbull contributed to technical support while doing bioinformatic developments and Corgi's training and maintenance. Heroen Verbruggen and Vanessa R. Marcelino contributed to workflow conceptualization of ChloroScan. All authors contributed to manuscript writing and revision.

# Figures and Tables

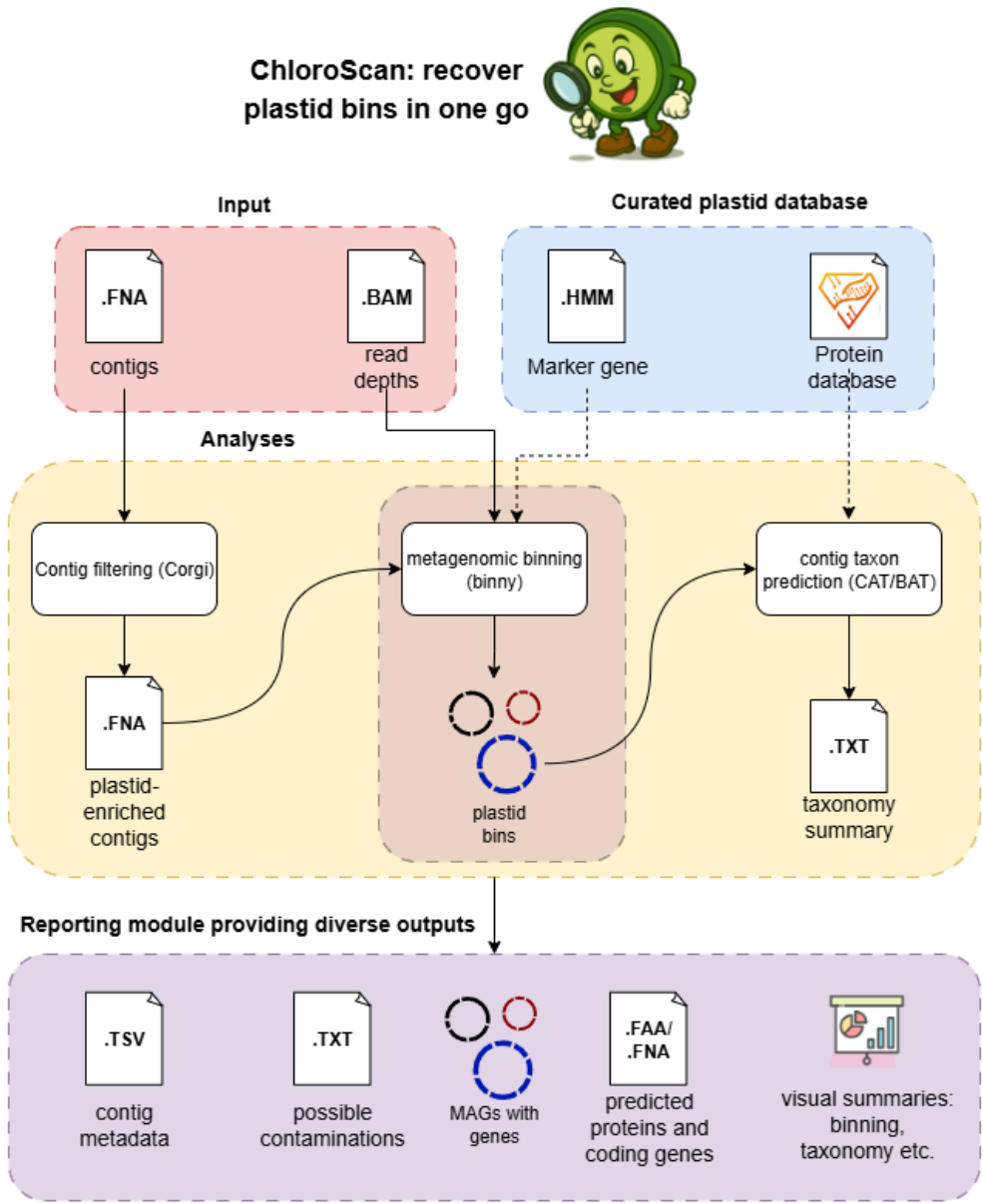

**Figure 1.** ChloroScan's workflow structure. ChloroScan contains the following modules: contig prediction by Corgi [50], binning by binny [47], taxonomic prediction by CAT/BAT [51], and a summary module that generates user-friendly information to investigate the contig and bin data, including plots to investigate bin homogeneity, a table with contig metadata and the predicted genes and proteins from MAGs. It takes assemblies in FASTA format and sorted BAM files as raw inputs. It is configured by a custom marker gene database in CheckM's format [21] and a protein database that works for binning and taxonomic predictions, respectively. Normal arrows represent passing files as inputs for steps and dotted lines represent passing configuration files or directories to the steps. The binning module (enclosed by a shaded box) is further engineered to target plastid genomes rather than prokaryotes.

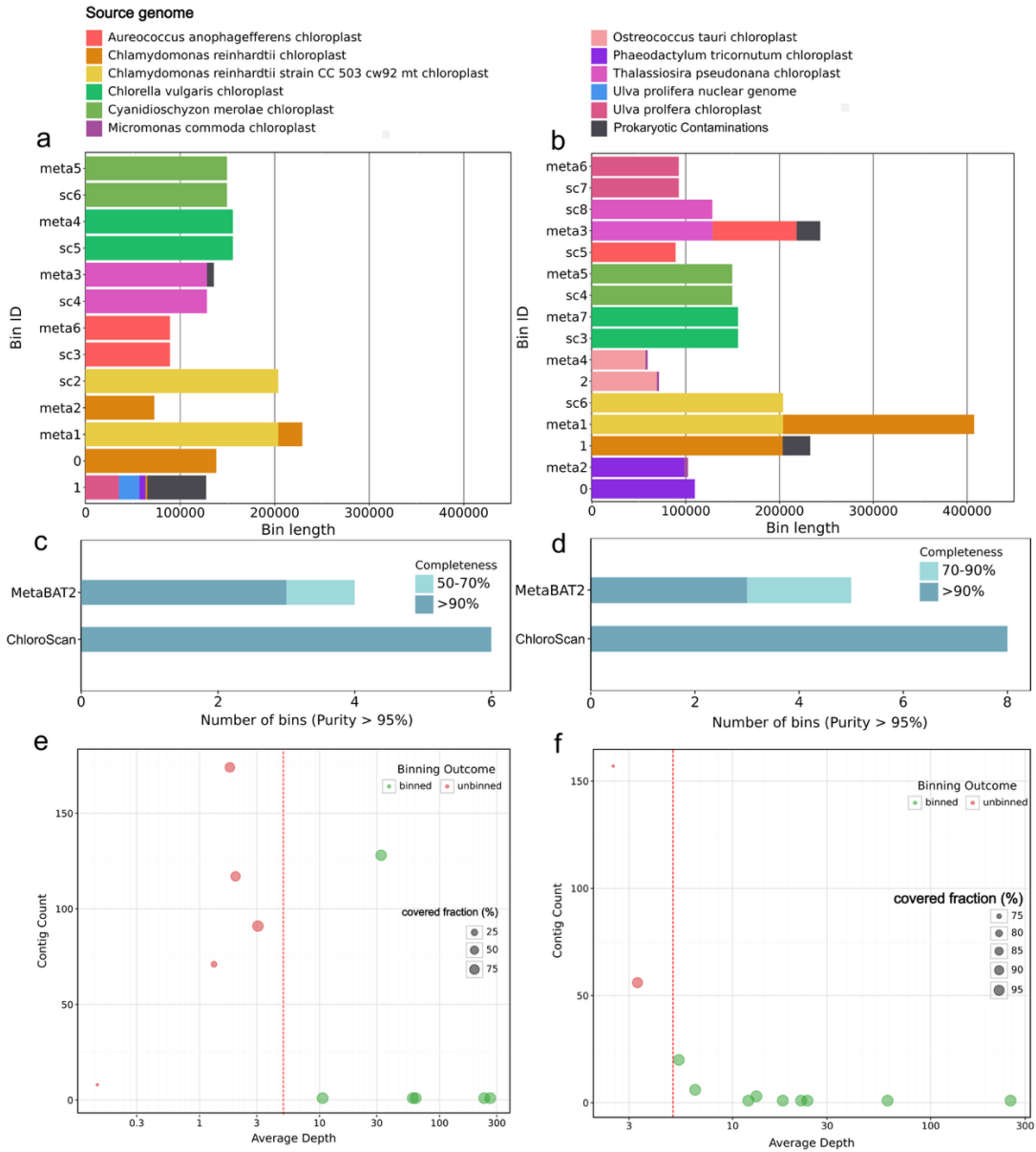

**Figure 2.** Plots of ChloroScan results show its effectiveness compared to MetaBAT2. The bar charts compare bins from ChloroScan to bins from MetaBAT2 in (a) single sample metagenome 1 and **(b)** single sample metagenome 2, in terms of their homogeneity by showing how many taxa are included in one bin. ChloroScan bins are labelled as digits and MetaBAT2 bins are prepended with "meta". Single contig MAGs recovered by ChloroScan are labelled as "sc". Prokaryotic contigs regardless of their taxa are labelled dark grey. The stacked bar charts on the right side demonstrate the count of ptMAGs in different quality classes from each tool in the

**(c)** sample 1 and **(d)** sample 2. The relationship between the average read depth calculated by binny's workflow, the level of fragmentations of each source genomes corresponding and their recovery status is plotted for sample 1 **(e)** and sample 2 **(f)**. The red dashed vertical lines represent an average depth of 5×.

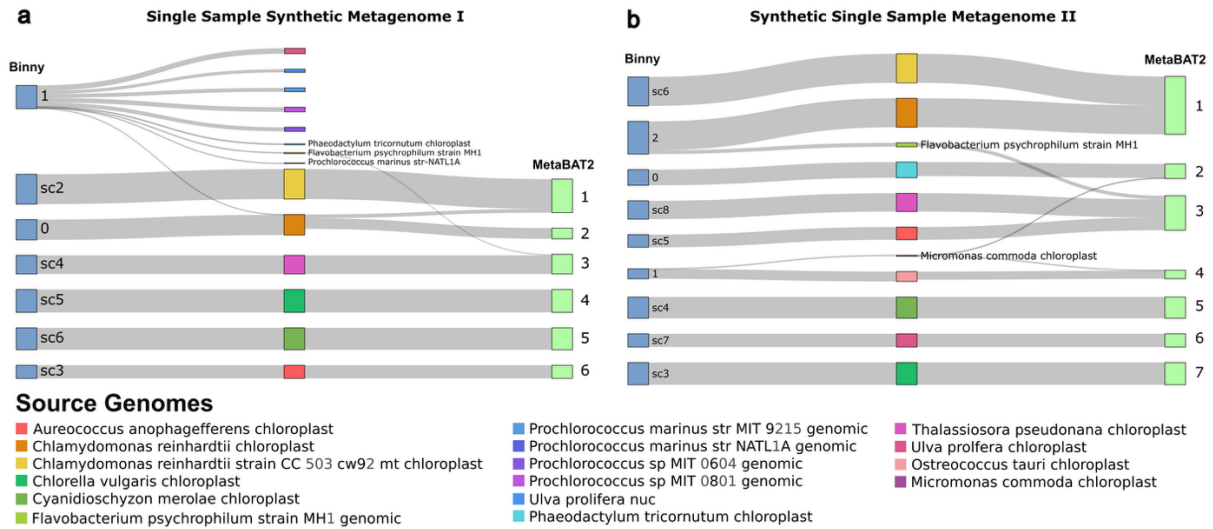

**Figure 3.** Mapping information from each bin to source genomes in the **(a)** synthetic single-sample metagenome 1 and **(b)** 2 based on the contig mapping information generated from CAMISIM. Grey bar widths refer to the percentage of source genome length taken by contigs. Here one contig has only one source genome mapped. Source genomes with too short contigs (colors invisible in the Figure) in the sample have their names labelled near the bar. Binny-recovered single-contig MAGs are prepended with "sc".

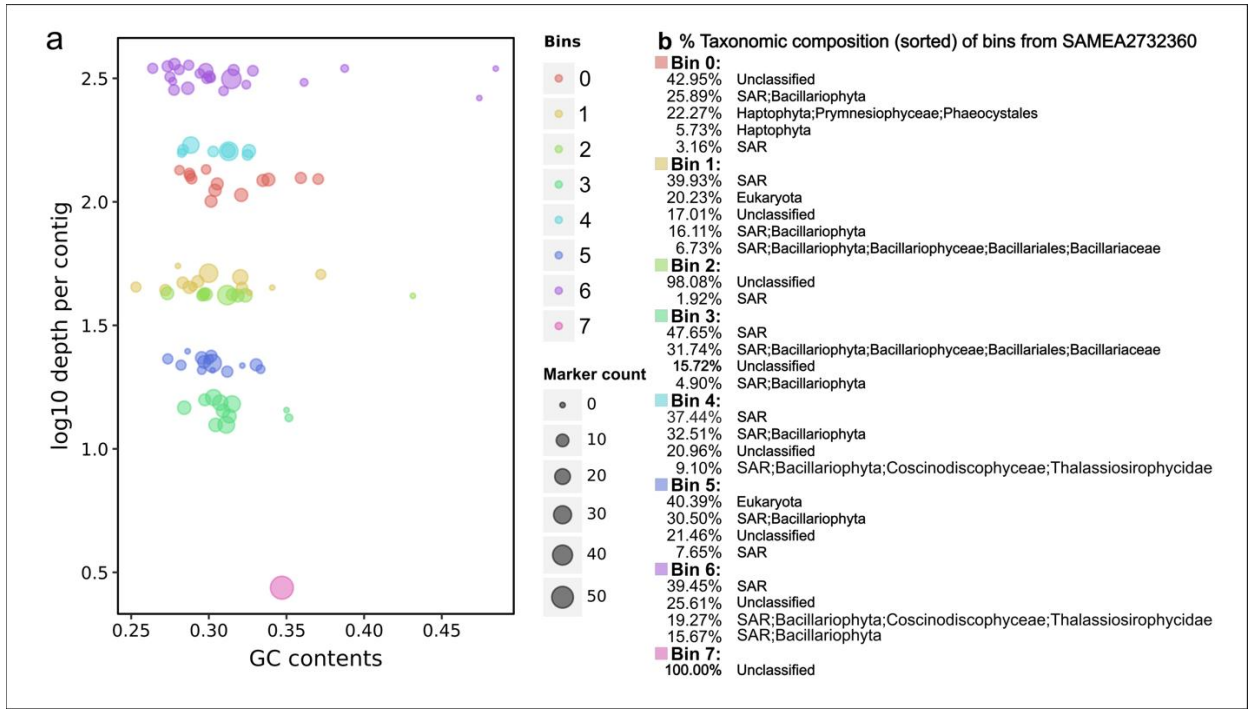

**Figure 4. Metagenome-assembled genomes from real marine metagenomes**. **a.** The GC x log10 average read depth plots of the sample SAMEA2732360. Marker gene count per contig is scaled by the dot size. **b.** Contig-level taxonomy composition of six bins from the sample SAMEA2732360 inferred by CAT. The sorted percentages of MAG length taken by each taxon are listed on the left side, and the detailed taxon lineages (adapted from NCBI taxonomy) for eukaryotic contigs are on the right side. The "Unclassified" refers to all categories without an exact taxon name (i.e.: root, unclassified, cellular organisms). The plot for SAMEA2189670 is in supplementary material (Figure S4).